\documentclass[twocolumn,showkeys,showpacs,preprintnumbers,prd,superscriptaddress,nofootinbib,shortlabels]{revtex4-1}
\usepackage{verbatim}
\usepackage[T1]{fontenc}
\usepackage[utf8]{inputenc}
\usepackage[american]{babel}
\usepackage{epsfig}
\usepackage{booktabs}
\usepackage{multirow}
\usepackage{dcolumn}
\usepackage{amsmath}
\usepackage{geometry}
\usepackage{mathtools}
\usepackage{amsfonts}
\usepackage{amssymb}
\usepackage{ulem}
\usepackage{epstopdf}
\usepackage{bm}
\usepackage{siunitx}
\usepackage{braket}
\usepackage{enumitem}
\usepackage{soul}
\usepackage[table]{xcolor}
\usepackage{color}
\usepackage{transparent}
\usepackage{pifont}
\usepackage{enumitem}
\usepackage{tikz}
\usetikzlibrary{shapes.geometric, arrows}
\definecolor{darkblue}{rgb}{0.0, 0.0, 0.55}
\definecolor{darkred}{rgb}{0.55, 0.0, 0.0}

\usepackage{hyperref}
\hypersetup{
    colorlinks=true, 
    linkcolor=darkblue,
    citecolor=darkblue,
    urlcolor=darkblue}

\makeatletter\let\expandableinput\@@input\makeatother

\tikzstyle{if} = [diamond, aspect=2, minimum width=1cm, minimum height=0.2cm, text centered, draw=black, fill=green!30]

\tikzstyle{arrow} = [thick,->,>=stealth]

\tikzstyle{startstop} = [rectangle, rounded corners, minimum width=1.5cm, minimum height=1cm,text centered, draw=black, fill=red!30]

\tikzstyle{process} = [rectangle, minimum width=3cm, minimum height=1cm, text centered, draw=black, fill=orange!30]

\begin{document}

\title{Observational bounds on Dark Matter–Admixed Neutron Stars from Gravitational Wave Data}

\author{Rafael M. Santos}
\email{rafael.mancini@inpe.br}
\affiliation{Divis\~ao de Astrof\'isica, Instituto Nacional de Pesquisas Espaciais, Avenida dos Astronautas 1758, S\~ao Jos\'e dos Campos, 12227-010, SP, Brazil}

\author{Rafael C. Nunes}
\email{rafadcnunes@gmail.com}
\affiliation{Instituto de F\'{i}sica, Universidade Federal do Rio Grande do Sul, 91501-970 Porto Alegre RS, Brazil}
\affiliation{Divis\~ao de Astrof\'isica, Instituto Nacional de Pesquisas Espaciais, Avenida dos Astronautas 1758, S\~ao Jos\'e dos Campos, 12227-010, SP, Brazil}

\author{Jaziel G. Coelho}
\email{jaziel.coelho@ufes.br}
\affiliation{Núcleo de Astrofísica e Cosmologia (Cosmo-Ufes) \& Departamento de Física, Universidade Federal do Espírito Santo, 29075-910 Vitória, ES, Brazil}
\affiliation{Divis\~ao de Astrof\'isica, Instituto Nacional de Pesquisas Espaciais, Avenida dos Astronautas 1758, S\~ao Jos\'e dos Campos, 12227-010, SP, Brazil}

\author{Jose C. N. de Araujo}
\email{jcarlos.dearaujo@inpe.br}
\affiliation{Divis\~ao de Astrof\'isica, Instituto Nacional de Pesquisas Espaciais, Avenida dos Astronautas 1758, S\~ao Jos\'e dos Campos, 12227-010, SP, Brazil}

\begin{abstract}
Recent gravitational-wave (GW) observations offer a unique opportunity to probe the fundamental nature of compact objects. A growing body of research has focused on exploring the role of dark matter (DM) through the concept of DM-admixed neutron stars (NSs), where the presence of DM can significantly alter key physical properties of NSs, such as their mass, radius, and tidal deformability, ultimately affecting the predicted GW waveform emitted during binary coalescences. In this work, we present a novel observational test that, for the first time, uses GW inspiral waveforms in terms of DM parameters to place constraints on the influence of DM inside NSs using real GW data. By reanalyzing signals from events such as GW230529, GW200115, and GW200105, we derive new upper bounds on the DM fraction, $F_{\chi}$, and particle mass, $m_{\chi}$, under the assumption that DM is described by a scalar field with a self-interaction potential. We find that the upper bound on $F_{\chi}$ depends on the specific binary system under analysis, indicating that different DM configurations can be consistent with observations in different ways. In particular, the event GW190814 may be compatible with a DM halo configuration. In contrast, the other events analyzed (GW230529, GW200105 and GW200115) are consistent with DM forming a core inside the NS, yielding strong upper bounds on $F_{\chi}$. The corresponding values for the mass scale $m_{\chi}$ are also discussed in the text. This work offers a new approach to probing DM in the context of compact NS objects through GW observations.
\end{abstract}



\maketitle

\section{Introduction}
\label{sec:introduction}

The term dark matter (DM) refers to a form of non-luminous matter that appears to dominate the mass-energy content of the universe. Its existence is inferred from a wide range of cosmological and astrophysical observations accumulated over recent decades, such as galaxy rotation curves, gravitational lensing, large-scale structure formation, and cosmic microwave background anisotropies (see \cite{Balazs:2024uyj,Cirelli:2024ssz} for a review). Despite its pervasive gravitational influence, DM has yet to be directly detected via electromagnetic interactions, and its fundamental nature remains one of the most pressing open questions in modern physics.

While the large-scale phenomenology of DM has been extensively tested in cosmology and galactic dynamics, it is natural to expect that some fraction of dark matter may also accumulate within compact stellar objects. In this context, neutron stars (NSs) emerge as exceptional natural laboratories to probe the physics of matter under extreme conditions. They host the densest known form of matter in the universe outside of black holes, with core densities exceeding nuclear saturation density. The internal structure and composition of NSs are governed by the still-unknown equation of state (EoS) of nuclear matter, which encapsulates the relationship between pressure, energy density, and other thermodynamic variables (see \cite{Rezzolla:2018jee,Lattimer:2000nx} for a comprehensive review). Given their extreme densities and long lifespans, neutron stars provide promising environments to test scenarios involving dark matter interactions and accumulation. Several studies have investigated the potential consequences of DM on the structure and evolution of NSs \cite{Ciarcelluti:2010ji,Kain:2021hpk,Leung:2011zz,Kumar:2025ytm,DelPopolo:2020hel,Li:2012ii,Yang:2024uzb,Scordino:2024ehe,Konstantinou:2024ynd,Das:2020vng,Liu:2024rix,Buras-Stubbs:2024don,Sun:2023cqr,Cronin:2023xzc,Liu:2023ecz,Arvikar:2025hej,Scordino:2024ehe,Liu:2023ecz,Parmar:2023zlg,Shirke:2023ktu,Kumar:2025cro,Vikiaris:2024jvq,Diedrichs:2023trk,Ruter:2023uzc,Hippert:2022snq,Cassing:2022tnn,Emma:2022xjs,Leung:2022wcf,Gleason:2022eeg,Lee:2021yyn,Thakur:2025zhi,Xiang:2013xwa}. One well-motivated scenario involves the accretion of DM by neutron stars over time, leading to configurations where DM is present in the core or coexists with ordinary nuclear matter in a mixed phase. In particular, some models predict the presence of mirror or asymmetric dark matter inside NS interiors, a possibility supported by known accretion mechanisms from the galactic halo \cite{Routaray:2024fcq,Sandin:2008db,Ivanytskyi:2019wxd,Rutherford:2024uix,Deliyergiyev:2023uer,Ciancarella:2020msu,Hippert:2021fch,Miao:2022rqj,Sarkar:2020tbv,Gresham:2018rqo}. Alternatively, DM may form an extended halo surrounding the neutron star, rather than penetrating its interior \cite{Ellis:2018bkr,Shawqi:2024jmk,Liu:2024qbe}. In both cases, the gravitational and possible non-gravitational interactions of DM can significantly alter the mass-radius relation, tidal deformability, cooling rates, and stability of neutron stars. These modifications can lead to observable astrophysical signatures \cite{Grippa:2024ach}.

The advent of gravitational wave (GW) astronomy has inaugurated a new era in the study of compact objects, offering unprecedented opportunities to probe the fundamental physics governing their structure and composition. A landmark moment occurred with the discovery of GW170817 event by the LIGO and Virgo collaborations. This multi-messenger event provided a unique and powerful probe of the EoS of dense nuclear matter, as it enabled simultaneous constraints from both gravitational and electromagnetic signals \cite{LIGOScientific:2017ync,LIGOScientific:2018cki, Malik:2018zcf, Radice:2017lry}. Since then, additional detections of compact binary coalescences involving neutron stars—either in neutron star-neutron star (NS-NS) or neutron star-black hole (NS-BH) systems—have been reported, further enriching the observational landscape \cite{Abac_2024,Abbott_2021_NSBH, Abbott_2020, Abbott_2021_NSBH, Abbott_2023}. These observations have made it possible to place tighter constraints on the properties of neutron stars. Importantly, they also offer a new avenue to test for the presence of exotic physics, such as DM within or around neutron stars \cite{Kumar:2025yei,Liu:2025cwy,Karkevandi:2021ygv,Das:2020ecp,Das:2021yny,Sen:2021wev,Das:2021wku,Mukherjee:2025omu,Bezares:2019jcb}.

In parallel, the era of precision X-ray astronomy has also contributed significantly to our understanding of neutron star interiors. Instruments such as NICER and XMM-Newton have enabled highly accurate measurements of neutron star masses and radii through pulse-profile modeling and spectral analysis. These measurements provide complementary constraints to those obtained from GW observations and are particularly valuable for probing the EoS at supranuclear densities \cite{Miller:2021qha,Riley:2021pdl}. Crucially, both gravitational wave and X-ray observations have been leveraged in theoretical studies exploring the possibility that neutron stars may contain, or be affected by, dark matter \cite{Shakeri:2022dwg, Routaray:2023spb}. The inclusion of DM, either as a component accumulated in the core or in an extended halo—can alter the star’s internal structure, cooling behavior, and tidal response. As such, compact object observations now serve as promising laboratories for constraining a broad range of DM models, from weakly interacting massive particles (WIMPs) to asymmetric and self-interacting dark matter candidates \cite{Karkevandi_2022, Grippa:2024ach, Ciancarella:2020msu, Kain_2021, Kouvaris:2007ay, Shakeri:2022dwg,Ellis:2018bkr,Barbat:2024yvi, Arvikar:2025hej}.

This work presents the first event-by-event reanalysis of gravitational-wave inspiral data in which dark-matter micro-parameters $(m_\chi, \lambda_\chi)$ and the DM fraction $F_\chi$ are directly sampled and propagated through a two-fluid TOV mapping into waveform parameters. This strategy enables direct constraints on DM microphysics from GW inspiral waveforms of real events. To achieve this, we analyze signals from confirmed compact binary coalescence events detected by the LIGO/Virgo/KAGRA (LVK) collaborations. We employ state-of-the-art Bayesian inference techniques to explore the full parameter space of binary systems, focusing on scenarios in which one or both neutron stars contain a non-negligible fraction of DM. In particular, we implement a parametric framework tailored to a theoretically well-motivated case: bosonic dark matter. \textit{To the best of our knowledge, this is the first study to place statistical constraints on the DM content within NSs by reanalyzing GWs events such as GW230529, GW200115, and GW200105. Our methodology combines real observational data with a rigorous Bayesian framework, enabling a systematic exploration of the impact of DM on the structure of compact objects. The results presented here open a novel and promising avenue for probing the microscopic properties of dark matter through astrophysical observations of neutron star binaries.}

The paper is organized as follows. In Section~\ref{theory}, we review the theoretical framework that motivates this study, introducing the main assumptions and presenting the key equations required for modeling the system. Section~\ref{data} describes the statistical methodology employed, along with the observational data sets used in our analysis. In Section~\ref{results}, we present and discuss our main results, highlighting their implications in the context of dark matter–admixed neutron stars. Finally, Section~\ref{final} summarizes our conclusions and outlines possible directions for future work.

\section{Dark matter admixed neutron stars}
\label{theory}

\begin{figure*}
    \centering
    \includegraphics[scale=0.25]{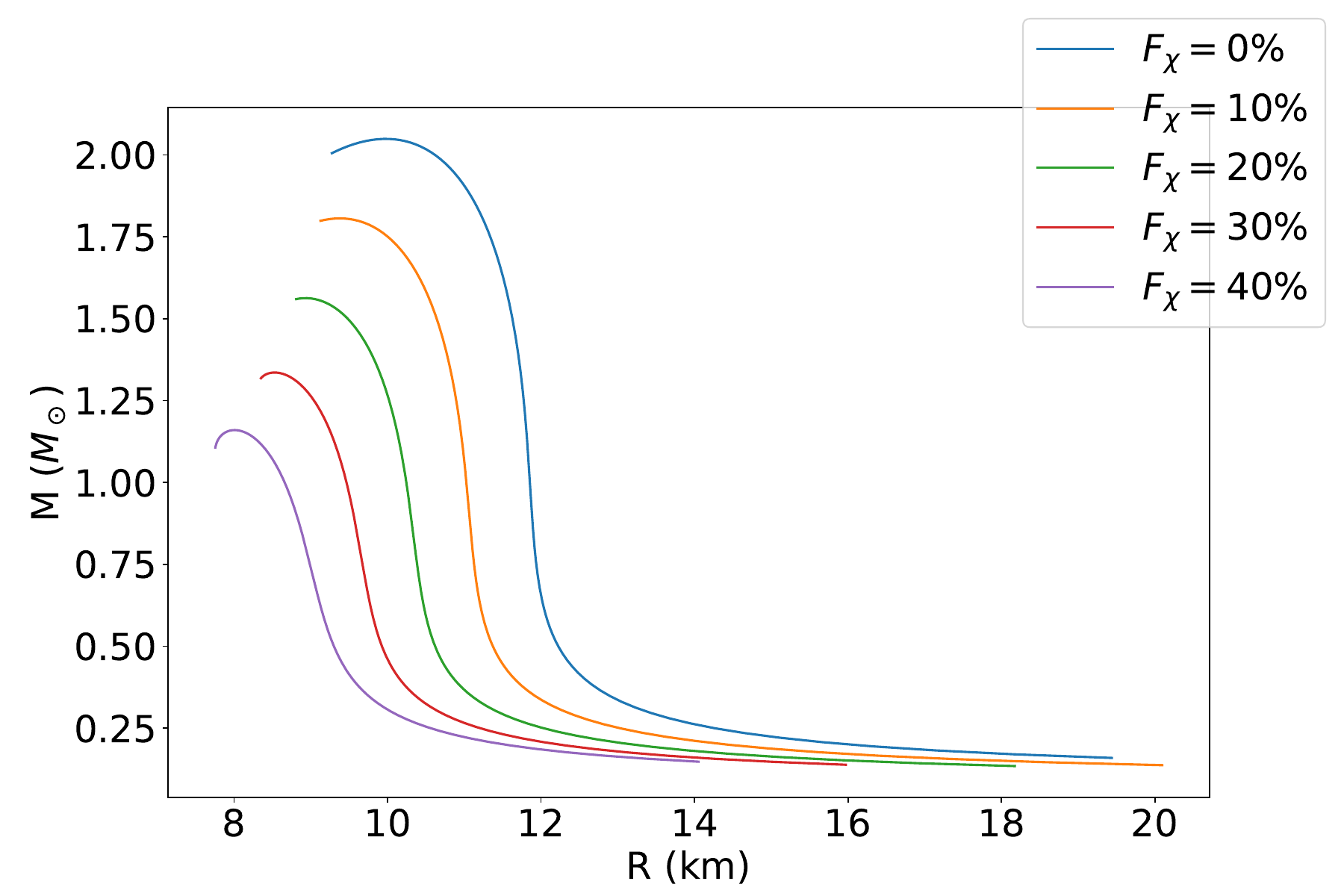} \,\,\,\,
    \includegraphics[scale=0.25]{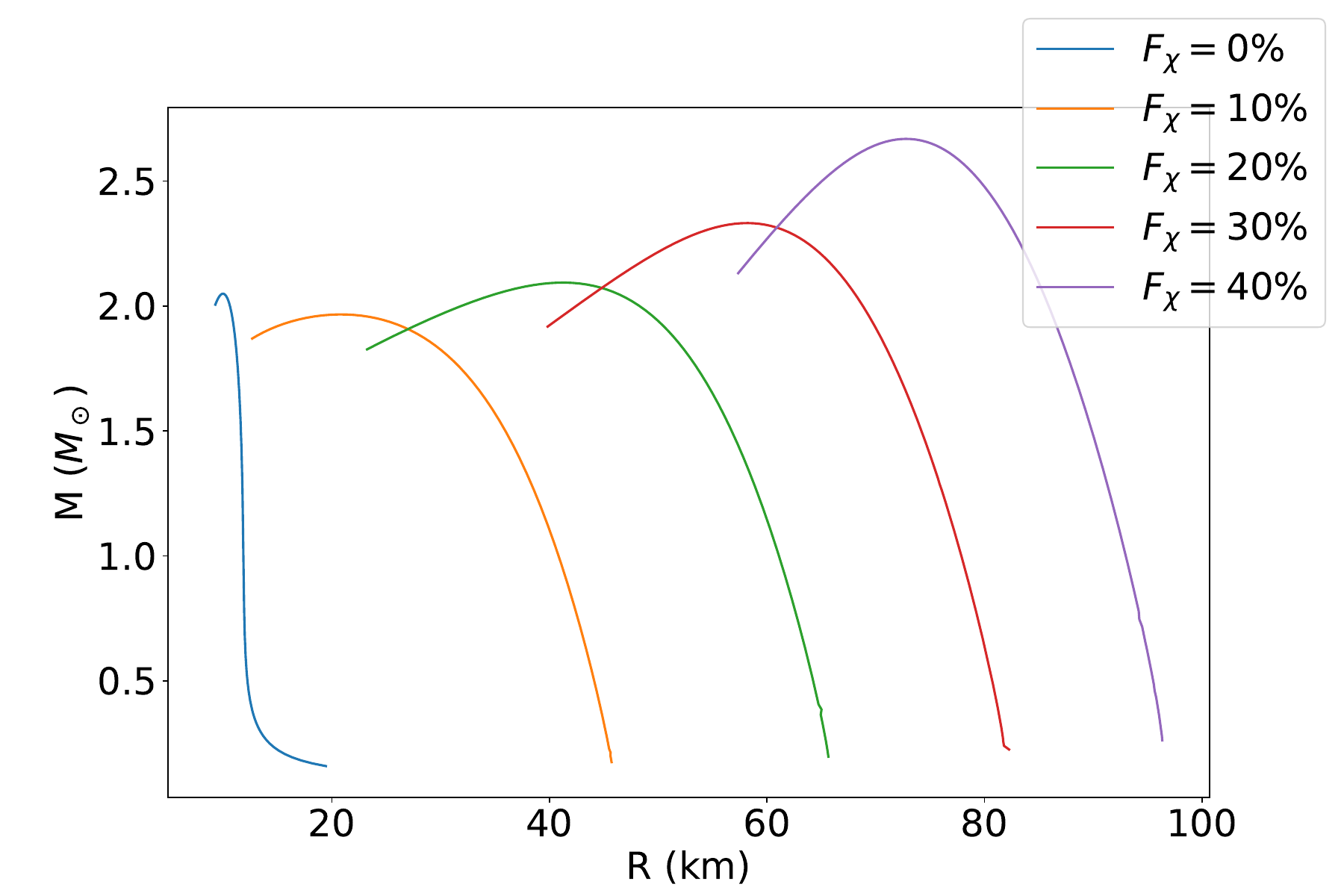}
    \caption{Mass–radius relations for DM-admixed configurations corresponding to cases (i) and (ii) discussed in Section~\ref{theory}. \textbf{Left panel:} scalar field parameters set to $m_\chi = 400~\mathrm{MeV}$ and $\lambda_\chi = \pi$. \textbf{Right panel:} scalar field parameters set to $m_\chi = 120~\mathrm{MeV}$ and $\lambda_\chi = \pi$.}
    \label{fig:MR400MeV}
\end{figure*}

In this work, we adopt a bosonic DM model, whose impact on the structure of NSs and the generation of GWs was thoroughly investigated in \cite{Karkevandi_2022} (see also \cite{Grippa:2024ach}). Due to the absence of degeneracy pressure in bosonic matter, it is essential to include self-interactions to prevent gravitational collapse. In this framework, DM particles are modeled as a complex scalar field $\phi$ with a quartic self-interaction potential given by $V(\phi) = (\lambda_\chi/4)\, |\phi|^4$, where $\lambda_\chi$ is a dimensionless coupling constant. Assuming that the DM component exists at zero temperature, the system can be described as a Bose-Einstein condensate \cite{Li:2012qf}, leading to an effective equation of state (EoS) given by \cite{Colpi_1986}
\begin{equation}
    p = \frac{m^4_\chi}{9\lambda_\chi}\left(\sqrt{1+\frac{3\lambda_\chi}{m^4_\chi}\epsilon} - 1 \right)^2,
\end{equation}
where $m_\chi$ is the mass of the DM particle.

In the context of compact bodies, spacetime can be described by the static, spherically symmetric metric given by
\begin{equation}
ds^2 = -e^{\nu(r)}dt^2 + \frac{dr^2}{1-2m(r)/r} +r^2d\Omega,  
\end{equation}
where $\nu(r)$ is the metric function, $m(r)$ is the mass enclosed within radius $r$ and $d\Omega^2 = d\theta^2 +\sin^2\theta d\phi^2$.
To model NSs admixed with DM, we assume an energy-momentum tensor composed of Baryonic Matter BM and DM fluids, $T^{\mu\nu} = T^{\mu\nu}_{\text{DM}} + T^{\mu\nu}_{\text{BM}}$, with only gravitational interaction between them. Both components are considered to be a perfect fluid, i.e.
\begin{equation}
    T^{\mu\nu}_{I} = (\epsilon_I + p_I)u^\mu u^\nu + p_I\eta^{\mu\nu},
\end{equation}
where $I=\text{DM,BM}$ stands for the DM and BM components, respectively. In this case, each energy-momentum tensor is conserved separately, giving rise to a set of equations of motion known as the two-fluid formalism of Tolman-Oppenheimer-Volkoff (TOV) \cite{tov_ref, tov_ref2, Kain_2021}.
\begin{align} \label{eq:2TOV_1}
    \frac{dp_{I}}{dr} &= - (p_{I} + \epsilon_{I}) \frac{m(r) + 4\pi r^3p}{r[r - 2m(r)]} \\ \label{eq:2TOV_3}
    \frac{dm}{dr} &= 4\pi r^2\epsilon,
\end{align}
where $m(r) = m_{\text{DM}}(r) + m_{\text{BM}}(r)$, $p=p_{\text{DM}} + p_{\text{BM}}$ and $\epsilon=\epsilon_{\text{DM}} + \epsilon_{\text{BM}}$. The solution to the equations above describes the hydrostatic equilibrium configuration of neutron stars admixed with dark matter (DANS), characterizing the distribution of DM within the star. These configurations fall into three general scenarios:
\begin{enumerate}[(i)]
    \item DM forms a core inside the NS. In this case, the radius of the DM component, $R_{\text{DM}}$, is smaller than the radius of the baryonic (baryonic matter, BM) component, $R_{\text{BM}}$, i.e., $R_{\text{DM}} < R_{\text{BM}}$.
    \item DM forms a halo around the NS, i.e., $R_{\text{DM}} > R_{\text{BM}}$.
    \item DM is distributed throughout the entire star, i.e., $R_{\text{DM}} = R_{\text{BM}}$.
\end{enumerate}

\begin{figure}
    \centering
    \includegraphics[width=1.1\linewidth]{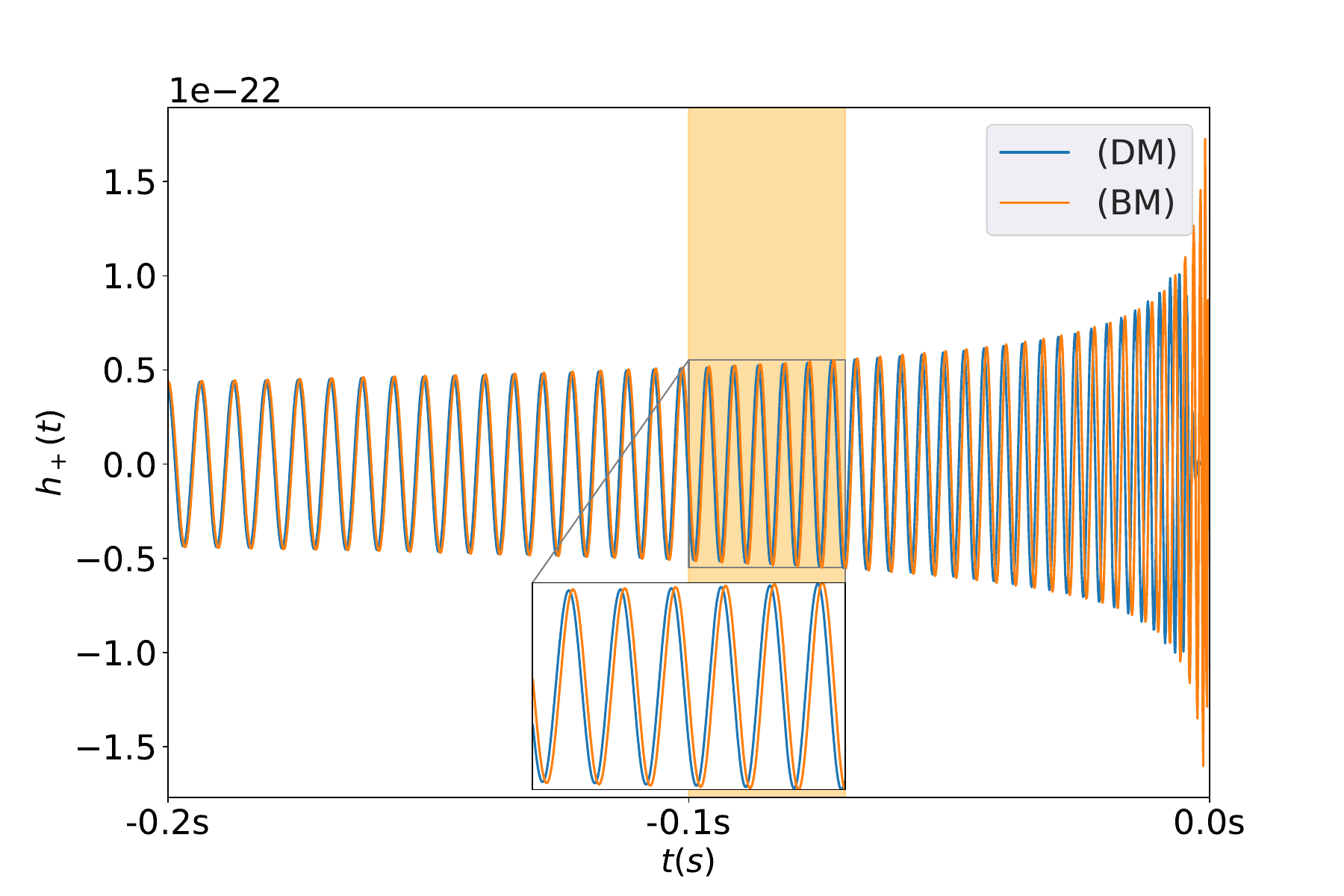}
    \caption{Simulated polarization signals during the inspiral phase are shown for two scenarios: a neutron star containing dark matter (DM curve) and one composed solely of baryonic matter (BM curve). The waveforms were computed using the \texttt{IMRPhenomNSBH} model, assuming fixed values of $F_\chi = 50\%$, $m_\chi \approx 220~\mathrm{MeV}$ and $\lambda_\chi = \pi$. All remaining intrinsic and extrinsic parameters were set according to those inferred for the GW230529 event.}
    \label{fig:DMWaveform}
\end{figure}

It is important to emphasize that even in scenarios (i) and (ii), DM and BM are still gravitationally coupled throughout the star. However, configurations featuring a pure DM core would require non-gravitational interactions between DM and BM \cite{Zhang:2020pfh}.

To solve the two-fluid TOV equations, boundary conditions and EoSs $p_I = p_I(\epsilon_I)$ must be chosen for each  matter component. Given central energy densities $e^c_{\text{BM}}$ and $e^c_{\text{DM}}$, we approximate the masses of each component near the center of the star, $m_{\text{BM}}(r\approx0)$ and $m_{\text{DM}}(r\approx0)$, through a Taylor expansion of the two-fluid TOV equations and calculate the central pressures, $p^c_{\text{BM}}(e^c_{\text{BM}})$ and $p^c_{\text{DM}}(e^c_{\text{DM}})$, with each matter component's EoS. We then perform a numerical integration of Eqs. \ref{eq:2TOV_1} and \ref{eq:2TOV_3} outwards up to a radius where one component's pressure vanishes. The integration then continues until the pressure of the other component vanishes, which marks the surface radius $R=R_{\text{BM}}$ or $R=R_{\text{DM}}$, depending on the DM configuration.

The total mass of the object is defined as:
\begin{equation}
M_T = M_{\text{BM}} + M_{\text{DM}},
\end{equation}
where $M_{\text{BM}} = m_{\text{BM}}(R_{\text{BM}})$ and $M_{\text{DM}} = m_{\text{DM}}(R_{\text{DM}})$ are the total enclosed masses of the BM and DM, respectively. To parameterize the amount of DM inside an NS, we adopt the following definition:
\begin{equation}\label{eq:DMFraction}
F_{\chi} = \frac{M_{\text{DM}}}{M_T}
\end{equation}
where $F_{\chi}$ is the fraction of DM inside the NS. Although $F_{\chi}$ is considered as an input parameter for the model, it can only be calculated once a solution has been found. Hence, we employ an optimization procedure to control the DM fraction of a solution: we use \texttt{gsl\_root\_fsolver\_brent}\cite{brent2013algorithms,gough2009gnu} to find the zero of the function $F_{\chi}(\epsilon^c_{\text{BM}}, \epsilon^c_{\text{DM}})-F^\text{in}_\chi$ for a fixed $\epsilon^c_{\text{BM}}$, where $F_{\chi}(\epsilon^c_{\text{BM}}, \epsilon^c_{\text{DM}})$ is a routine that solves the two-fluid TOV equations and calculates the DM fraction from the boundary conditions, and $F^\text{in}_\chi$ is the input DM fraction. 

Figure~\ref{fig:MR400MeV} displays the mass-radius relations for the DM-admixed configurations corresponding to cases (i) and (ii) defined above, shown in the left and right panels, respectively. The DM parameters adopted are $m_\chi = 400\,\mathrm{MeV}$ (left panel) and $m_\chi = 120\,\mathrm{MeV}$ (right panel), with the same coupling constant $\lambda_\chi = \pi$ in both cases. As previously noted in \cite{Karkevandi_2022, Grippa_2024ach}, configurations with a DM core tend to correlate a reduced maximum mass with increasing DM fraction, while the opposite behavior is observed for DM halo configurations. Moreover, DM halos can have radii extending significantly beyond the typical range of NS radii \cite{Karkevandi_2022}, although only the BM radius is directly accessible via electromagnetic observations. 
These results highlight that markedly different astrophysical configurations can lead to significant modifications in the mass-radius relation of neutron stars.

On the other hand, the primary impact of DM on gravitational wave signals from compact binary coalescence (CBC) systems is manifested through the tidal deformability, $\Lambda$, of a DANS. This quantity characterizes the extent to which a neutron star deforms in response to the tidal gravitational field of its companion. The dimensionless tidal deformability is defined as \cite{Ellis:2018bkr}
\begin{equation}
    \Lambda = \frac{2}{3}k_2C^{-5},
\end{equation}
where $C=M_T/R$ is the compactness parameter and $k_2$ is the $\ell =2$ tidal Love number \cite{Damour_2009, Flanagan_2008}, which can be computed from the following expression \cite{Hinderer:2007mb}:
\begin{align} \label{eq:lovenumber}
    k_2 = &\frac{8C^5}{5}(1-2C)^2[2+2C(y_R-1)-y_R]\times   \\ \nonumber
    &\times\{2C[6-3y_R+3C(5y_R-8)]+ \\ \nonumber
    &+4C^3[13-11y_R+C(3y_R-2)+ \\ &+2C^2(1+y_R)]  \nonumber
    3(1-2C)^2\times \\ \nonumber
    &\times[2-y_R + 2C(y_R-1)]\ln(1-2C)\}^{-1},
\end{align}
where $y_R=y(r=R)$ is the solution to the differential equation
\begin{equation} \label{eq:k2diffeq}
    r\frac{dy(r)}{dr} +y(r)^2 + y(r)F(r)+r^2Q(r)=0,
\end{equation}
in which
\begin{align}
    F(r) &= \frac{r-4\pi r^3(\epsilon - p)}{r-2m(r)}, \\
    Q(r) &= \frac{4\pi r[5\epsilon+9p+\sum_I \frac{(e+p)}{\partial p_I/\partial \epsilon_I}-\frac{6}{4\pi r^2}]}{r-2m(r)} - \\ \nonumber
    &-4\left[\frac{m(r)+4\pi r^3 p}{r^2(2m(r)/r)} \right]^2.
\end{align}
Note that the functions $F(r)$ and $Q(r)$ couple Eq.~(\ref{eq:k2diffeq}) with Eqs.~(\ref{eq:2TOV_1})--(\ref{eq:2TOV_3}), and therefore must be solved simultaneously in order to compute the tidal deformability $\Lambda$.

In the context of gravitational waves, $\Lambda$ contributes a phase correction to the waveform at fifth post-Newtonian (PN) order, relative to the Newtonian term, and also influences the coalescence time: neutron stars with higher tidal deformability experience stronger tidal interactions and thus merge earlier. Considering NS-NS coalescing systems, the leading order tidal contribution arises from a combination of each component's tidal deformability, given by
\begin{equation}
    \tilde{\Lambda} = \frac{16}{13}\left[ \frac{(m_1 + 12m_2)m_1^4 \Lambda_1 + (m_2 + 12m_1)m_2^4\Lambda_2}{(m_1 + m_2)^5} \right],    
\end{equation}
where $m_i$ and $\Lambda_i$ are the mass and tidal deformability of the ith component, respectively. However, it must be emphasized that for NS-BH systems, which are the focus of this work, $\tilde{\Lambda}$ is proportional to the tidal deformability of the corresponding NS. That is, in all of our main analyses, we set $\Lambda_1 = 0$.

Figure~\ref{fig:DMWaveform} presents a comparison between simulated gravitational waveforms from a NS–BH merger system, considering both a DANS and a pure BM neutron star. The intrinsic and extrinsic parameters used to generate these waveforms follow those of the event GW230529. For the DM-admixed case, we adopt $F_\chi = 50\%$, $m_\chi \approx 220\,\mathrm{MeV}$, and $\lambda_\chi = \pi$. These parameters were selected to maximize the deviation from the pure BM scenario (with $\Lambda \sim 5000$), corresponding to a configuration in which the DM forms a halo around the NS.

It is important to note that the difference between the two waveforms is barely perceptible during the inspiral phase, with an average phase shift of approximately $0.002$ radians. The distinction becomes significant only in the final moments before merger, reaching a maximum phase shift of about $\sim\pi/6$ radians roughly $0.0002$ seconds before coalescence. Consequently, constraints on DM parameters cannot be derived from tidal deformability effects alone, given the current sensitivity of GW detectors such as LVK. This limitation motivates the alternative analysis strategy adopted in this work, which will be discussed in the next section.

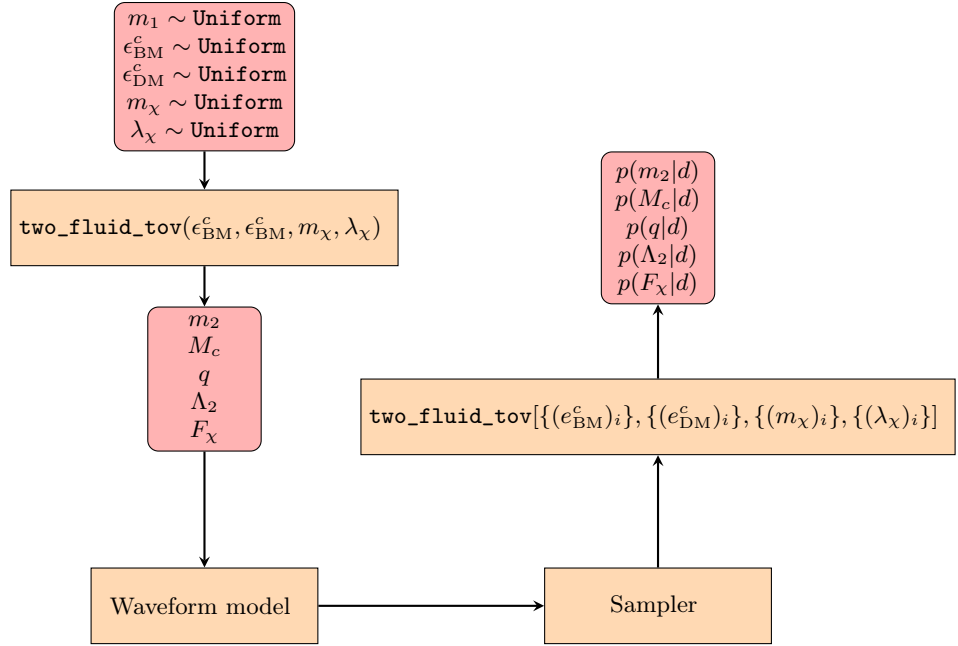
\begin{figure*}
    \centering
\begin{tikzpicture}[every text node part/.style={align=center}]
    \node (start) [startstop] {
    $m_1 \sim \texttt{Uniform}$ \\
    $\epsilon^c_{\text{BM}} \sim \texttt{Uniform}$ \\
    $\epsilon^c_{\text{DM}} \sim \texttt{Uniform}$ \\
    $m_\chi \sim \texttt{Uniform}$ \\
    $\lambda_\chi \sim \texttt{Uniform}$
    };
    \node (proc1) [process, below of=start, yshift=-1cm] {
    $\texttt{two\_fluid\_tov}(\epsilon^c_{\text{BM}}, \epsilon^c_{\text{BM}},m_\chi,\lambda_\chi)$
    };
    \node (start2) [startstop, below of=proc1, yshift=-1cm] {
    $m_2$ \\
    $M_c$ \\
    $q$ \\
    $\Lambda_2$ \\
    $F_\chi$
    };
    \node (proc2) [process, below of=start2, yshift=-2cm] {
    Waveform model
    };
    \node (proc3) [process, right of=proc2, xshift=5cm] {
    Sampler
    };

    \node (proc4) [process, above of=proc3, yshift=1.5cm] {
    $\texttt{two\_fluid\_tov}[\{(e^c_{\rm BM})_i\}, \{(e^c_{\rm DM})_i\}, \{(m_\chi)_i\}, \{(\lambda_\chi)_i\}]$
    };
    \node (start3) [startstop, above of=proc4, yshift=1.5cm] {
    $p(m_2|d)$ \\
    $p(M_c|d)$ \\
    $p(q|d)$ \\
    $p(\Lambda_2|d)$ \\
    $p(F_\chi|d)$
    };

    \draw [arrow] (start) -- (proc1);
    \draw [arrow] (proc1) -- (start2);
    \draw [arrow] (start2) -- (proc2);
    \draw [arrow] (proc2) -- (proc3);
    \draw [arrow] (proc3) -- (proc4);
    \draw [arrow] (proc4) -- (start3);
  
\end{tikzpicture}
    \caption{Flowchart illustrating the statistical strategy employed in this work to integrate the two-fluid TOV formalism with the EoS parameters, enabling the generation of waveforms required for a robust analysis of gravitational wave event signals. For a parameter $\theta$, we write its marginal posterior as $p(\theta|d)$, where $d$ is the data, and the set of samples drawn from it as $\{(\theta)_i\}$. This approach ensures that the relevant physical parameters associated with the dark matter sector are properly incorporated into the sampling procedure.}
    \label{fig:priorflow}
\end{figure*}

\section{Methodology and Dataset}
\label{data}

Our analyses are based on Bayesian parameter estimation and focus on  NS-BH coalescence events. In particular, we consider the events GW230529, GW200115, GW200105, and GW190814. The latter, GW190814, is especially intriguing, as the inferred mass of its secondary component exceeds the commonly accepted upper limit for the maximum neutron star mass, given the constraints from NICER data for PSR J0740+6620 \cite{Miller:2021qha, Riley:2021pdl} ($2.14^{+0.10}_{-0.09} \text{ M}_\odot$), thereby opening the possibility that the secondary object could be a DANS. For the analysis carried out in this work, we assume that the secondary component of GW190814 is a NS.

In Bayesian parameter estimation, given a model $\mathcal{M}$ characterized by a set of parameters $\theta$, the posterior probability distribution of $\theta$ conditioned on the strain data $d$ is given by

\begin{equation}
    p(\theta|d,\mathcal{M}) = \frac{\mathcal{L}(d|\theta, \mathcal{M}) \pi(\theta|\mathcal{M})}{\mathcal{Z}},
\end{equation}
where $\mathcal{L}(d|\theta,M)$ is the likelihood function, $\pi(\theta|M)$ is the prior distribution, and  
\begin{equation}
    \mathcal{Z} = \int \mathcal{L}(d|\theta,\mathcal{M}) \pi(\theta|\mathcal{M}) \, d\theta
\end{equation}
is the evidence. We employ a standard Gaussian noise likelihood given by 
\begin{equation}
\label{eq:L}
\mathcal{L}(d|\theta,\mathcal{M}) \propto \exp{\left[-\frac{1}{2}\sum_I \langle h_I(\theta) - d_I | h_I(\theta) - d_I \rangle \right]},
\end{equation}
where $I$ denotes the detector label. The noise-weighted inner product $\langle \cdot | \cdot \rangle$ is defined as  
\begin{equation}
    \langle a|b \rangle \equiv 4\,\mathrm{Re}\int_{f_{\rm min}}^{f_{\rm max}} \frac{a^*(f) b(f)}{S_{I,n}(f)} \, df,
\end{equation}
where $f_\text{min}$ and $f_\text{max}$ are the cutoff frequencies, $S_{I,n}(f)$ is the noise power spectral density of detector $I$, obtained from the LVK posterior sample releases \cite{Abac_2024,Abbott_2023}, and $a^*(f)$ is the complex conjugate of $a(f)$.

The GW signal from an aligned-spin CBC system depends on four intrinsic parameters: the primary and secondary masses $m_1$ and $m_2$, and the dimensionless spin parameters $\chi_1$ and $\chi_2$. However, the chirp mass $M_c=(m_1 m_2)^{3/5} / (m_1 + m_2)^{1/5}$ and mass ratio $q=m_2/m_1$ are usually the ones preferred to be sampled rather than the component masses, since the $M_c$ provides the dominant contribution to the GW signal, and thus is better constrained. Moreover, the parameters $M_c$ and $q$ can then be used to obtain $m_1$ and $m_2$ during the sampling process of a Bayesian parameter estimation analysis of a GW signal.

To correctly model the detector response, seven extrinsic parameters must also be considered: the luminosity distance, $d_L$; the orbital inclination angle, $\iota$; the position of the sky, specified by the right ascension, $\alpha$ and the declination, $\delta$; the polarization angle, $\psi$; the phase at a reference frequency, $\phi_c$; and the coalescence time, $t_c$. Combining the intrinsic and extrinsic parameters, the full parameter space of a typical Bayesian analysis of a GW signal is given by

\begin{equation}
\label{eq:parameter_space1}
\theta \equiv\! \Bigl\{M_c, q, \chi_1, \chi_2, d_L, \alpha, \delta, \iota, \psi, t_c, \phi_c \Big\}.
\end{equation}

A common approach to parameter estimation with EoSs involves solving the TOV equations for a range of boundary conditions at each step of the sampling process. This enables the tidal deformability to be calculated from an interpolated $\Lambda(m, \textbf{E})$ curve \cite{Lackey:2014fwa, LIGOScientific:2018cki}, where $\textbf{E}$ is a vector containing the EoS parameters and $m$ is the NS source mass. The high computational cost of this approach can be partially mitigated through the use of interpolated likelihood tables \cite{HernandezVivanco:2020cyp}. However, in the current DANS scenario, this method encounters some problems. First, considering core DM configurations, the maximum mass decreases as $F_\chi$ increases. Meanwhile, the opposite happens with DM halo configurations. This poses a problem for the construction of an interpolated function $\Lambda(m,\textbf{E})$, since some EoS parameters cause a range of values for $m$ to be unavailable. Furthermore, the signal-to-noise ratio (SNR) of current data on NS-BH coalescing systems is too low to allow constraints to be put on the DM EoS parameters given only the effect of tidal deformability on GW generation.

In this work, we adopt a different strategy by directly sampling over the parameters $m_1$, $m_\chi$, $\lambda_\chi$, $\epsilon^c_{\text{BM}}$, and $\epsilon^c_{\text{DM}}$. These are then transformed into the physical parameters $M_c, q, F_\chi, \mbox{ and } \Lambda_2 $ by solving the two-fluid TOV equations, which are then used as inputs to the waveform source model to allow the correct incorporation of DM effects into waveform generation. The posterior distributions of the physical parameters $M_c, q, F_\chi \mbox{ and } \Lambda_2$ are recovered at the end of the sampling procedure. Figure~\ref{fig:priorflow} presents a flowchart illustrating this strategy. This approach circumvents the issues discussed in the previous paragraph, as it eliminates the need to construct an explicit function $\Lambda(m, \textbf{E})$ at each step of the sampling procedure —-- the tidal deformability is instead obtained directly from the TOV solution. Furthermore, the NS source mass contributes significantly to constraining the DM sector, often providing the dominant constraints. The parameter space in this approach can be summarized as follows:

\begin{align}
    \theta \equiv\! \Bigl\{& m_1, \chi_1, \chi_2, d_L, \alpha, \delta, \iota, \psi, t_c, \phi_c, \nonumber 
    m_\chi, \lambda_\chi, \\ &\epsilon^c_{\text{BM}}, \epsilon^c_\text{DM}, F_\chi\Big\}.
\end{align}

For all the events considered for this work, we use the waveform approximant \texttt{IMRPhenomNSBH}~\cite{Thompson_2020}, which incorporates a state-of-the-art phase model derived from the \texttt{NRTidal} approximants~\cite{Dietrich:2017aum, Dietrich:2018uni, Dietrich:2019kaq}. This approximant includes tidal disruption effects and is calibrated for aligned spin systems with magnitudes constrained to $|\chi_1| < 0.5$ and $|\chi_2| < 0.05$ for the primary and secondary components, respectively.

To study the correlation of BM EoSs with the DM parameters, we analyzed each event considering the well-established realistic EoSs SLy4 \cite{Chabanat:1997un} and APR4 \cite{Akmal:1998cf}, which support NS masses up to $2.06\text{ M}_\odot$ and $2.15\text{ M}_\odot$, respectively, in accordance with the lower bounds for maximum NS mass imposed by the analysis of PSR J0740+6620 \cite{Miller:2021qha, Riley:2021pdl}. For the event GW190814, due to the peculiar nature of its secondary component, we also considered the stiffer EoS MPA1 \cite{Muther:1987xaa}, which supports NS masses up to $2.5M_\odot$. It is important to note that the aforementioned EoSs are also aligned with the constraints on tidal deformability imposed by the LVK analysis of GW170817, $\Lambda_{1.4}\lesssim 580$, where $\Lambda_{1.4}$ is the tidal deformability of a $1.4\text{ M}_\odot$ NS \cite{LIGOScientific:2018cki,Malik:2018zcf}.  

We note that nuclear-theory uncertainties remain significant at supranuclear densities. In particular, chiral-EFT–based extrapolations and phenomenological constructions admit a range of soft to stiff EoSs and possible phase transitions \cite{Tews2018,Tews2019}. These uncertainties imply that DM constraints obtained from GW data are conditional on baryonic-matter modeling. In this work we use SLy4/APR4/MPA1 as representative EoSs and quantify sensitivity to these choices via prior tests; a full marginalization over nuclear EoS families (spectral or piecewise-polytropic) would propagate nuclear uncertainty into wider posterior regions for $F_\chi$ and is a natural extension of this analysis.

The priors for the intrinsic and extrinsic parameters follow the low-spin priors adopted by the LVK analyses for each respective event~\cite{Abac_2024,Abbott_2023}. For the additional DM parameters, we assume uniform priors. Specifically, we set $\epsilon^c_{\text{BM}} \sim \texttt{Uniform}(0.4, 1.5)$ and $\epsilon^c_{\text{DM}} \sim \texttt{Uniform}(0.0, 5.0)$, with both quantities expressed in units of~$\mathrm{GeV}/\mathrm{fm}^3$. Additionally, we adopt $m_\chi \sim \texttt{Uniform}(80, 1000)$ (in MeV), and $\lambda_\chi \sim \texttt{Uniform}(10^{-20}, 2\pi)$. These ranges were chosen to include the parameter space previously explored in~\cite{Karkevandi_2022}, allowing for a more meaningful comparison and interpretation of our results. For consistency checks, we conducted extensive tests of prior sensitivity, focusing in particular on the central energy densities $\epsilon^c_{\mathrm{BM}}$ and $\epsilon^c_{\mathrm{DM}}$. For $\epsilon^c_{\mathrm{BM}}$, we adopted the range 0.4–1.5 GeV/fm$^3$, which encompasses typical densities of standard NS EoS while excluding solutions with $\Lambda > 5000$, known to be incompatible with waveform models. We further verified that using a narrower prior (0.3–1.0 GeV/fm$^3$) yields more uniform $\Lambda$ distributions without significantly affecting mass constraints. For $\epsilon^c_{\mathrm{DM}}$, the broader range 0.0–5.0 GeV/fm$^3$ enables exploration of diverse DM configurations, although values above $\sim$2 GeV/fm$^3$ are likely unphysical for bosonic DM.

Our analyses were carried out using the \texttt{parallel-bilby}~\cite{Smith_2020} package in combination with the \texttt{dynesty}~\cite{Speagle_2020} nested sampler. The sampling settings were selected to strike a balance between computational cost and inference accuracy. In particular, we set the number of live points (\texttt{nlive}) to 1000, the number of autocorrelation times (\texttt{nact}) to 10, and employed the \texttt{acceptance-walk} strategy with 100 walks.


\section{Results}
\label{results}

\begin{figure*}
    \centering
    \includegraphics[width=0.42\linewidth]{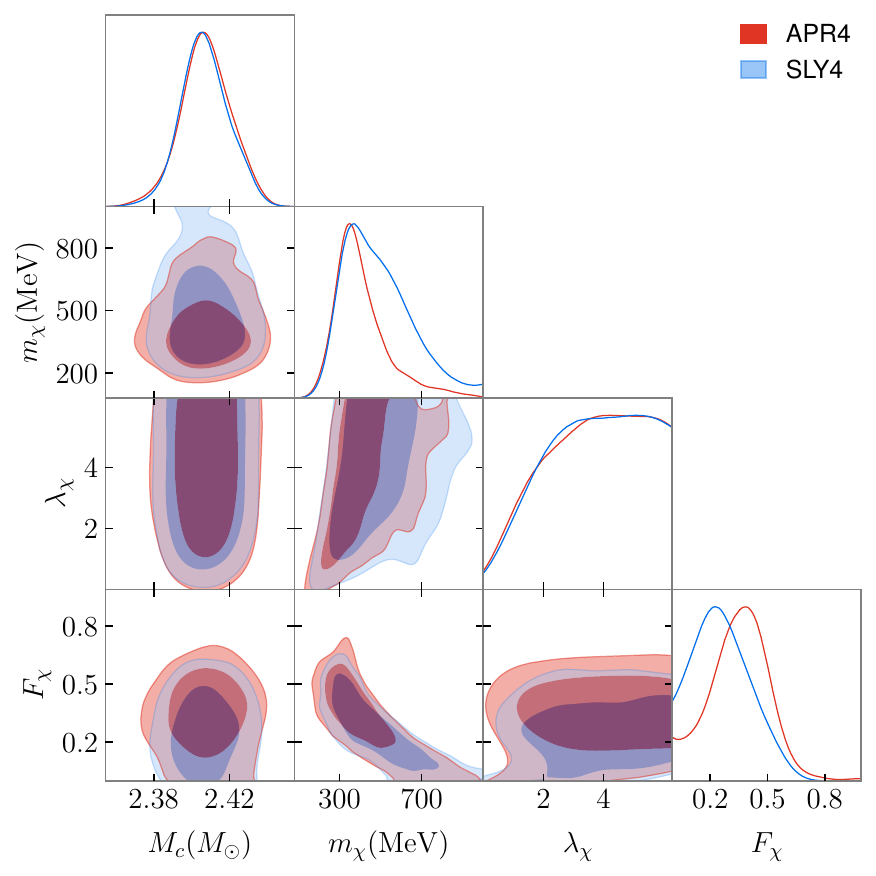}
    \includegraphics[width=0.42\linewidth]{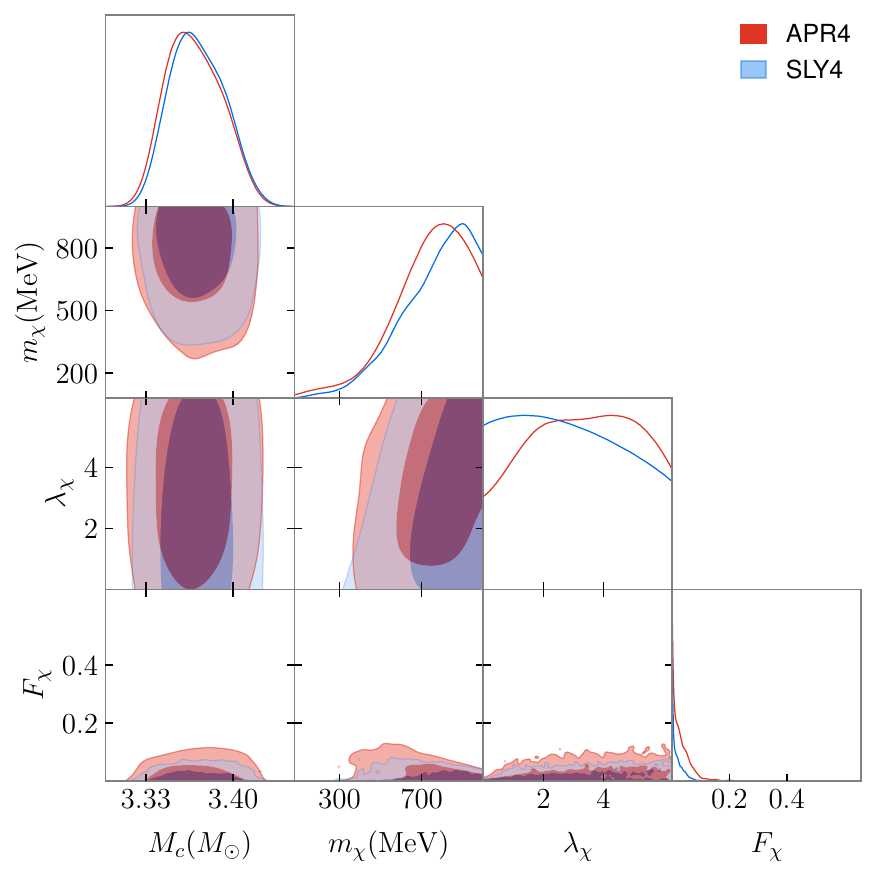}
    \caption{\textbf{Left panel:} Marginalized posterior distributions and 68\% and 95\% CL contours for source chirp mass and DM parameters $m_\chi$, $\lambda_\chi$ and $F_\chi$ for merger event GW200115 \cite{Abbott_2021_NSBH} considering the BM EoSs SLy4 \cite{Chabanat:1997un} and APR4 \cite{Akmal:1998cf} \textbf{Right panel:} Same as in left panel, but for event GW200105 \cite{Abbott_2021_NSBH}}
    \label{fig:GW20}
\end{figure*}

\begin{figure}
    \centering
    \includegraphics[width=0.9\linewidth]{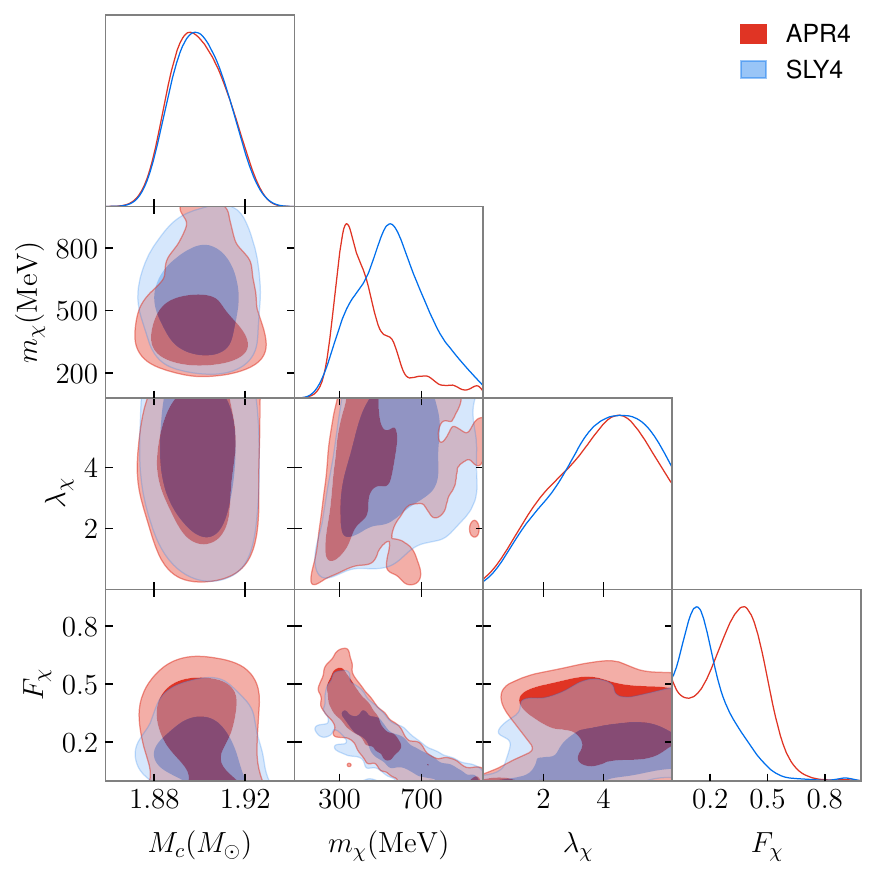}
    \caption{Same as Fig. \ref{fig:GW20}, but for merger event GW230529 \cite{Abac_2024}.}
    \label{fig:GW23}
\end{figure}
Our results are summarized in Figs. \ref{fig:GW20} and \ref{fig:GW23}, which show the  marginalized posteriors together with 68\% and 95\% confidence level (CL) contours for the source chirp mass, DM fraction and the scalar field's coupling constant and mass. Each event was analyzed with two well-established realistic EoSs for BM --- SLy4 and APR4. The event GW190814, due to its putative higher secondary component's mass, was further analyzed considering the BM EoS MPA1. For all events analyzed, constraints on the coupling constant $\lambda_\chi$ were loose, following an overall trend of $\lambda_\chi \gtrsim 1$. As shown in \cite{Karkevandi_2022}, the coupling constant $\lambda_\chi$ has a small correlation with mass and tidal deformability for DM core scenarios. Meanwhile, this correlation is strong for DM halo scenarios. Therefore, it is expected that GW events favoring DM halo scenarios yield better constraints for $\lambda_\chi$.
In what follows, we adopt the standard convention in which upper limits are reported at the 95\% CL, while all other parameter estimates are given at the 68\% CL.

We highlight that the choice of BM EoS had negligible impact on the inferred source chirp mass. However, there is a clear correlation between the choice of BM EoS and the inferred DM parameters --- particularly considering the DM fraction $F_\chi$. On this note, it is important to remark that for events GW230529, GW200115, and GW200105, the choice of BM EoS does not change the overall shape of the DM parameters' inferred posteriors. Rather, there are only minor differences in the bounds obtained, with the softer EoS SLy4 yielding the tighter constraints on the DM fraction $F_\chi$, a tendency that was already observed in \cite{Barbat:2024yvi,Arvikar:2025hej, Shakeri:2022dwg} in light of data from the NICER experiment and GW170817 for both bosonic and fermionic DM. For that reason, for the aforementioned GW events, we report the SLy4 constraints unless stated otherwise.

The analyses of events GW230529, GW200115, and GW200105 yielded constraints on $F_\chi$ that are compatible with a scenario without DM. In particular, the event GW200105 yielded the constraint $F_\chi < 6.1\%$ . Combined with its constraint on the scalar field's mass, $m_\chi > 435 \text{ MeV}$, this region of the parameter space is characterized by the DM core scenario, and hence justifies the lack of constraint for the coupling constant $\lambda_\chi$. The results of the analysis of GW200105 marginally agree with previous constraints obtained from comparisons with data from the NICER experiment and GW170817 \cite{Ellis:2018bkr,Grippa:2024ach,Barbat:2024yvi, Arvikar:2025hej, Shakeri:2022dwg}, which showed that $F_\chi \lesssim 5\%$ considering soft EoSs such as SLy4 and APR4.

\begin{figure}
    \centering
    \includegraphics[width=0.9\linewidth]{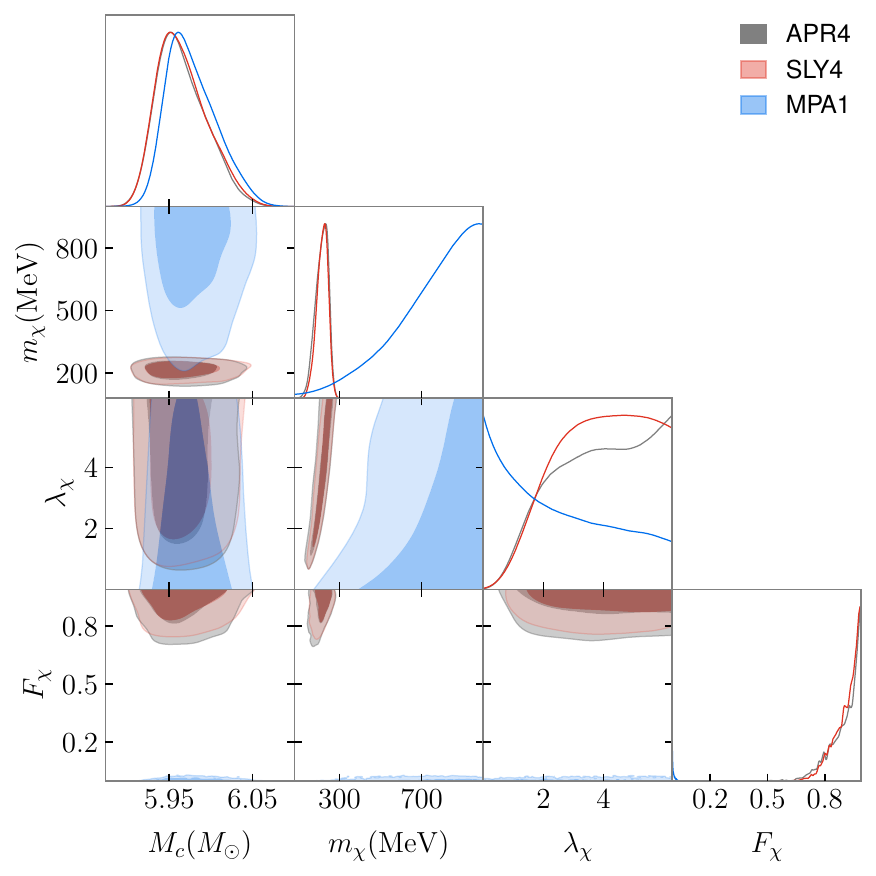}
    \caption{Marginalized posterior distributions for source chirp mass, and DM parameters $m_\chi$, $\lambda_\chi$ and $F_\chi$ for merger event GW190814 \cite{Abbott_2020} considering the BM EoSs SLy4 \cite{Chabanat:1997un}, APR4 \cite{Akmal:1998cf} and MPA1 \cite{Muther:1987xaa}}
    \label{fig:GW19}
\end{figure}

It should be noted that the events GW230529, GW200115 and GW200105 have secondary components with probabilities $\gtrsim 90\%$ of being neutron stars, mainly due to the source masses of their secondary components falling within the commonly accepted NS mass range \cite{Abbott_2021_NSBH, Abac_2024}. As a result, their expected masses are supported by the BM EoSs used in this work. Nevertheless, the results from events GW200115 and GW230529 showed similar support for high DM fractions. Specifically, the constraints were $F_\chi < 53.3\%$ for GW200115, while for GW230529 we obtained $F_\chi < 45.2\%$. Combined with their constraints on the DM mass, which were $m_\chi = 485^{+100}_{-200} $ MeV and $m_\chi = 558^{+200}_{-200}$ MeV for GW200115 and GW230529, respectively, these results also favor DM core configurations. These larger bounds can be an effect of systematic bias --- as will be further discussed, the choice of prior over the central energy densities can affect the inferred bounds on the $F_\chi$.

As can be seen in Fig \ref{fig:GW19}, the event GW190814 was an outlier among the events chosen for this work, since it was the event for which the choice of EoS made the most impact. From the LVK analysis, its secondary component's mass was estimated to be $m_2 = 2.59^{+0.08}_{-0.09} \, M_\odot$ at 90\% CL \cite{Abbott_2020}. This range of masses is not supported by the softer EoSs SLy4 and APR4 and, in this context, can only be achieved by DM halo configurations. Specifically, the results for this event yielded the constraints $F_\chi > 78.7\%$ and $m_\chi=216^{+30}_{-20}$ MeV (SLy4), the latter being the tightest constraint obtained for this parameter considering the events analyzed in this work. It is noteworthy to mention that these bounds, although extreme, align marginally with the tidal deformability constraint obtained from GW170817.

\begin{table}[h!]
\centering
\caption{ Summary of the constraints on \( m_\chi \) and \( F_\chi \) obtained for each GW event and BM EoS.}
\label{tab:gw_parameters}
\begin{tabular}{llcc}
\toprule
\textbf{Event} & \textbf{EoS} & \( m_\chi \) (MeV) & \( F_\chi \) \\
\midrule
\multirow{2}{*}{GW230529} 
  & SLy4 & $558^{+200}_{-200}$ & $< 0.452$ \vspace{1mm} \\
  & APR4 & $440^{+60}_{-200}$ & $< 0.556$ \\
\midrule
\multirow{2}{*}{GW200115} 
  & SLy4 & $485^{+100}_{-200}$ & $< 0.533$ \vspace{1mm} \\ 
  & APR4 & $408^{+70}_{-100}$ & $< 0.589$ \\
\midrule
\multirow{2}{*}{GW200105} 
  & SLy4 & $> 435$ & $< 0.0610$ \vspace{1mm} \\
  & APR4 & $> 403$ & $< 0.0952$ \\
\midrule
\multirow{3}{*}{GW190814} 
  & SLy4 & $216^{+30}_{-20}$ & $> 0.787$ \vspace{1mm} \\
  & APR4 & $214^{+40}_{-20}$ & $> 0.756$ \vspace{1mm} \\
  & MPA1 & $> 366$ & $< 0.022$ \\
\bottomrule
\end{tabular}
\end{table}

\begin{figure*}
    \centering
    \includegraphics[width=0.7\linewidth]{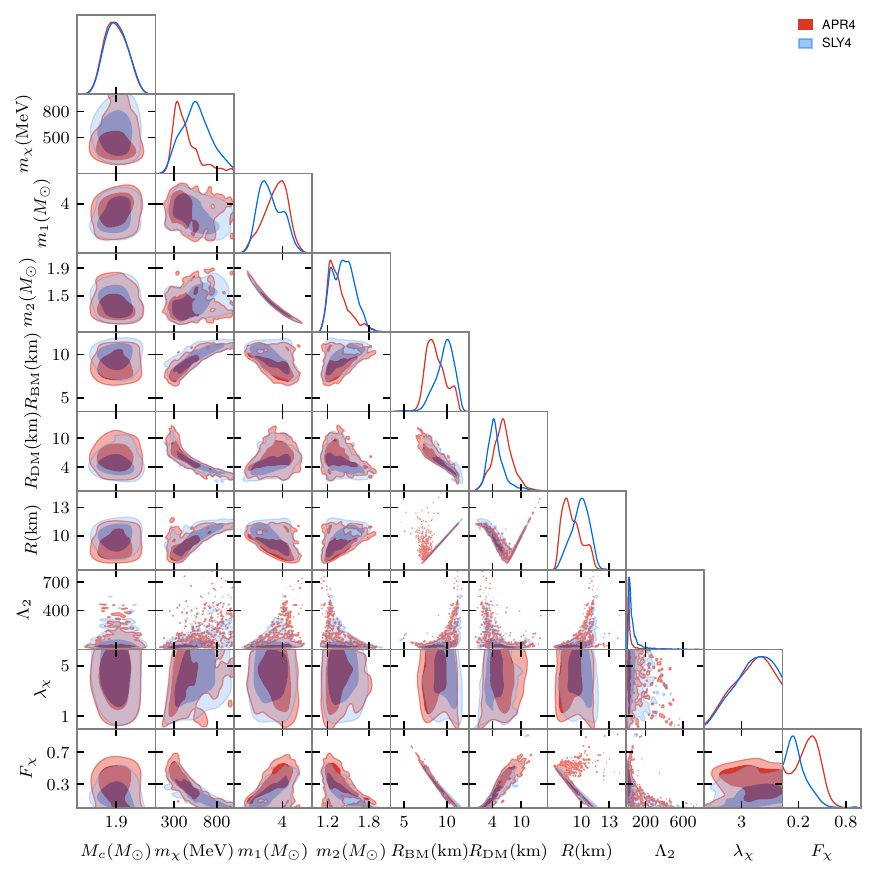}
    \caption{Same as Fig.~\ref{fig:GW23}, but also showing the baseline parameters: component source masses ($m_1$, $m_2$), baryonic radius $R_\text{BM}$, dark-matter radius $R_\text{DM}$, outermost radius $R$, and the tidal deformability of the secondary component $\Lambda_2$.}
    \label{fig:corner_gw230529}
\end{figure*}

\begin{figure*}
    \centering
    \includegraphics[width=0.7\linewidth]{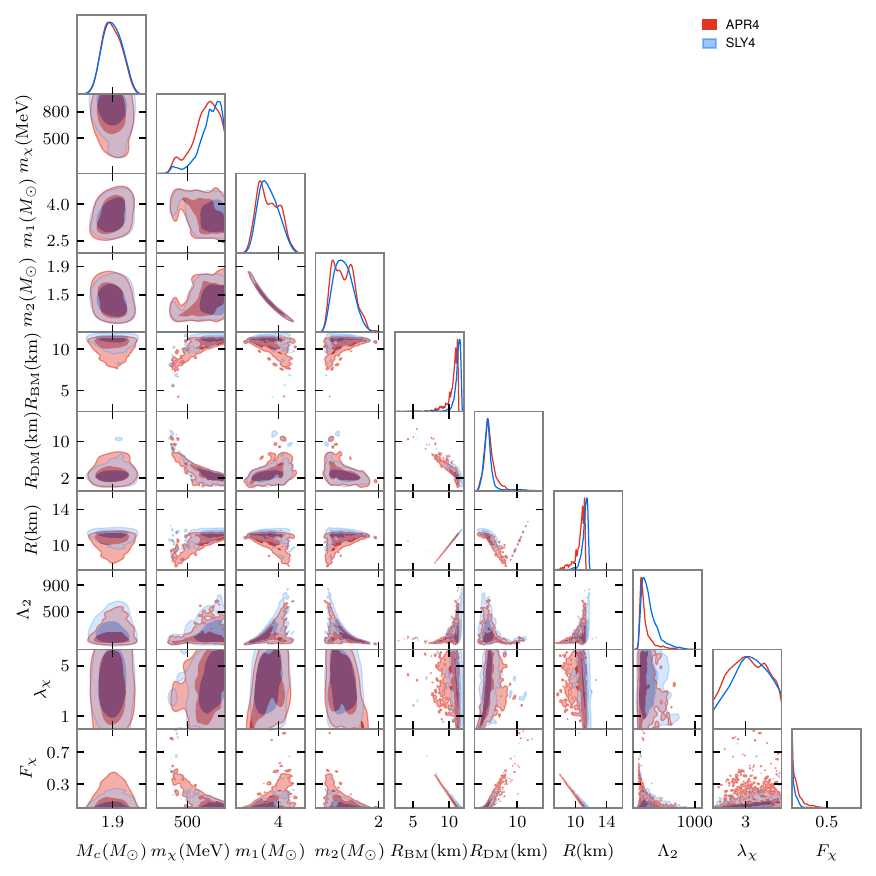}
    \caption{Same as Fig.~\ref{fig:corner_gw230529}, but adopting the prior $\epsilon^c_\text{BM}\sim \text{Uniform}(0.3,1.0)\ \mathrm{GeV/fm^3}$ for the BM central energy density. Under this prior choice, the bound tightens to $F_\chi < 20.5\%$ at 95\% CL.}
    \label{fig:prior_test}
\end{figure*}

The possibility that the secondary component of GW190814 is a DANS with DM core configuration has already been explored before by considering stiff, relativistic mean field BM EoSs models capable of achieving its inferred mass range \cite{Das:2021yny}. To explore further this possibility, we ran another analysis of event GW190814 using the stiffer BM EoS MPA1, which supports masses up to $\sim 2.5\text{ M}_\odot$. Fig. \ref{fig:GW19} shows the results of this run, which yielded the constraints $F_\chi < 2.2\%$ and $m_\chi > 366$ MeV (MPA1), supporting a DM core scenario for GW190814. Nonetheless, it is important to mention that EoSs capable of supporting high NS masses can be inconsistent with constraints obtained from tidal deformability and from heavy-ion collisions, which could put as a more likely scenario that of an extremely light black hole \cite{Fattoyev:2020cws}.
Future high-SNR observations could break this degeneracy by better constraining $\Lambda$ through tidal deformability measurements.

Table \ref{tab:gw_parameters} summarizes all the observational constraints obtained and discussed above, highlighting the events and equations of state used in each analysis. Figure~\ref{fig:corner_gw230529} shows the joint posterior for GW230529, 
highlighting the correlations between global structure variables ($R$, $\Lambda_2$), the component radii ($R_{\rm BM}$, $R_{\rm DM}$), and the DM sector ($F_\chi$, $\lambda_\chi$, $m_\chi$). The posterior favors core-like configurations (smaller $R_{\rm DM}\lesssim R_{\rm BM}$) for soft nuclear EoSs, while allowing halo-like solutions only at larger $m_\chi$ and $F_\chi$. The anti-correlation between $R$ (or $\Lambda_2$) and $F_\chi$ 
reflects the reduced compactness in halo-dominated models. To test robustness, we repeated the analysis with a narrower prior on the baryonic central energy density. Figure~\ref{fig:prior_test} summarizes the impact on ($\Lambda_2,R,R_{\rm BM},R_{\rm DM}$), demonstrating that while the numerical upper bounds on $F_\chi$ shift (through induced changes in $\Lambda_2$), the morphological preference (core vs halo) remains stable across EoSs.

Our baseline prior for the baryonic central energy density is
$\epsilon^c_\text{BM}\sim\texttt{Uniform}(0.4,1.5)\ \mathrm{GeV/fm^3}$, while for the dark sector we adopt $\epsilon^c_\text{DM} \sim \texttt{Uniform}(0.0,5.0)\ \mathrm{GeV/fm^3}$,
$m_\chi \sim \mathrm{Uniform}(80,1000)\ \mathrm{MeV}$, and $\lambda_\chi \sim \mathrm{Uniform}(10^{-20},2\pi)$, consistent with the ranges used in our main analyses.\footnote{See Sec.~III for the full prior specification and sampling settings.} Because the waveform model is only partially calibrated at very large tidal deformabilities, extremely small values of $\epsilon^c_\text{BM}$ can generate solutions with $\Lambda_2 \gtrsim 5000$. To assess the possible impact of such prior-driven effects on $\Lambda_2$ and $F_\chi$, we performed an additional run with
$\epsilon^c_\text{BM} \sim \mathrm{Uniform}(0.3,1.0)\ \mathrm{GeV/fm^3}$. This choice broadens support at lower densities and produces an approximately uniform prior-predictive distribution for $\Lambda_2$ in the range $0$–$5000$, thereby mitigating the tendency of the posterior to concentrate at very low $\Lambda_2$. The corresponding shift in inferred radii leads to modestly tighter upper bounds on $F_\chi$ (e.g., ${\lesssim}30\%$ at 95\% CL for GW230529), while the qualitative morphological preference (core-like vs.\ halo-like), assessed through the relation $R_{\rm DM}\lessgtr R_{\rm BM}$, remains unchanged within statistical uncertainties across the EoSs considered. These tests suggest that observational constraints on $F_\chi$ may be sensitive to the quality of the signal and to prior choices through $\Lambda_2$ modeling, although the qualitative core/halo inference appears robust under reasonable variations of $\epsilon^c_\text{BM}$.

One of the most notable and important events involving neutron stars is GW170817. In this work, following the new methodology developed here (summarized in Fig. \ref{fig:priorflow}), we note that analyzing a long-duration signal such as GW170817 entails a high computational cost, which can also lead to numerical stability challenges when evaluating the signal simultaneously from two neutron stars. Our code will soon be optimized and made publicly available\footnote{\url{https://github.com/rafamancin1/2fluidTOV_C}}, allowing the community to review and contribute to further developments. We also highlight a qualitative and theoretical analysis in this DM context that was recently presented in \cite{Grippa:2024sfu, Kumar:2024zzl, Mariani:2023wtv}. Clearly, a more robust implementation and a direct comparison with real signal data should be pursued in future studies.

\section{Conclusion}
\label{final}
In this work, we derived the first observational bounds in the literature on the possible properties of dark matter–admixed neutron stars (DANS) from gravitational-wave (GW) data. Our analysis is based on four NS–BH merger events: GW230529, GW200115, GW200105, and GW190814. We focused on constraining the dark matter fraction inside neutron stars and exploring the correlation between dark matter parameters and different baryonic matter (BM) equations of state (EoSs). To this end, we developed a dedicated parameter-estimation strategy that incorporates the DM sector directly into the sampling procedure while ensuring computational feasibility for real-event analyses.  

For the events GW230529, GW200115, and GW200105, we find that softer EoSs yield the most stringent bounds on the DM fraction, with GW200105 giving the strongest constraint, $F_{\chi} < 6.1\%$. In contrast, GW190814 shows the opposite behavior: the stiffest EoS produces the tightest bound, $F_{\chi} < 2.2\%$, while softer EoSs allow $F_{\chi} > 78.7\%$, a result incompatible with a purely baryonic neutron star.  

The statistical correlations shown in Figs.~\ref{fig:corner_gw230529} and~\ref{fig:prior_test} highlight two main trends: (1) larger $R_{\mathrm{BM}}$ values are associated with smaller $R_{\mathrm{DM}}$, consistent with expectations for DM core scenarios, and (2) the $\Lambda$ distribution peaks at relatively low values ($\Lambda \lesssim 80$). The latter trend is largely driven by our choice of prior on $\epsilon^c_{\mathrm{BM}}$, which naturally disfavors high-$\Lambda$ solutions. When adopting a narrower prior ($\epsilon^c_{\mathrm{BM}} \in [0.3,1.0]$ GeV/fm$^3$), the GW230529 analysis produces tighter constraints ($F_\chi < 20.5\%$ at 95\% CL, compared to the original 45.2\%), indicating that our reported bounds are conservative. Importantly, the qualitative preference for DM core solutions remains robust under different prior choices.

Our results also indicate that, for GW230529, GW200115, and GW200105, the data favor a DM-core configuration, whereas GW190814 is more consistent with a DM-halo scenario, unless a stiff EoS is assumed. This behavior is largely driven by the total system mass, which plays a dominant role in parameter-estimation convergence, especially given the relatively low signal-to-noise ratios (SNRs) of the available data. These findings represent the first direct GW-based constraints on DM properties in neutron stars and open a new observational window for testing dark matter in the context of compact objects. The detection of future gravitational-wave events with higher SNRs will play a crucial role in improving our understanding of the dark matter contribution to the internal structure of neutron stars, particularly in the transition region between the baryonic and dark matter components. Higher-quality data will enhance sensitivity to subtle effects on the tidal deformability and mass–radius relation, enabling more precise constraints on the DM fraction and its spatial distribution inside neutron stars. 

Building upon our robustness tests in Sec.~\ref{results}, our study highlights an important methodological finding: the absolute upper bounds on the DM fraction, $F_{\chi}$, exhibit a non-negligible dependence on the physical choice for the baryonic central energy density, $\epsilon_{\rm BM}^{\rm c}$. This is particularly evident for events like GW230529, where a narrower prior on $\epsilon_{\rm BM}^{\rm c}$ leads to a tighter constraint ($F_{\chi} < 20.5\%$ compared to $45.2\%$). This physical prior-sensitivity arises because $\epsilon_{\rm BM}^{\rm c}$ directly influences the tidal deformability $\Lambda$, which is weakly constrained by the low-SNR data available for these events. However, we emphasize that the qualitative morphological conclusion (i.e., the preference for a DM core configuration over a halo scenario for GW230529, GW200115, and GW200105) proves to be robust under the explored variations of the $\epsilon_{\rm BM}^{\rm c}$ prior. This robustness underscores that while the precise observational bounds may be sensitive to the choice of baryonic physics, the overarching inference regarding the spatial distribution of DM inside these NSs is a stable result of our analysis.

A comprehensive approach would simultaneously sample flexible parametrizations of the baryonic EoS (spectral or piecewise-polytropic) and the dark sector. Imposing physical hyper-priors (causality, thermodynamic stability and $M_{\max}\gtrsim2\,M_\odot)$ would allow the propagation of nuclear-theory uncertainty into $F_\chi$ constraints. We endorse this approach as a natural next step and note that the present prior-sensitivity tests indicate the qualitative impact expected from such marginalization.

It is worth noting that a promising future direction would be to combine information from multiple events for a population-level inference analysis (see \cite{LIGOScientific:2025pvj}). For instance, one could fix the DM mass and coupling strength uniformly and attempt to infer the DM fractions, $F_\chi$, by combining analyses of all available GW data. Currently, there are not yet enough detected events containing neutron stars for this methodology to be applied. Nevertheless, population inference—combining data from several events to constrain shared physical parameters—represents a novel and potentially insightful approach to the study of DM in NS systems. We plan to pursue this direction in future work, which will require the development of a dedicated analysis strategy.

Our analysis indicates that the DM core scenario cannot be statistically excluded. However, given the current data quality, observational constraints on DM sector parameters may still depend on prior choices and theoretical assumptions entering each analysis. While the DM core interpretation is favored for most events, this conclusion shows only modest sensitivity to the choice of priors. In particular, the $\epsilon^c_{\mathrm{BM}}$ prior range influences the $\Lambda$ distribution and, consequently, the bounds on $F_\chi$, although the overall core/halo morphological preference remains stable. In conclusion, our results and the methodology developed here indicate that further advances in computational modeling, together with the detection of new events with higher signal quality, are still required to study the DM sector using neutron stars in the context of GW astronomy.

\section*{Data Availability}
The datasets used in this research are publicly available on \url{https://gwosc.org/}. The source codes used in this research will be publicly available soon at \url{https://github.com/rafamancin1/2fluidTOV_C}

\begin{acknowledgments}
We thank the anonymous referee for their valuable suggestions and comments, which have helped to improve this work. R.M.S. thanks CNPq (141351/2023-3) for partial financial support. R.C.N thanks the financial support from the Conselho Nacional de Desenvolvimento Cient\'{i}fico e Tecnologico (CNPq, National Council for Scientific and Technological Development) under the project No. 304306/2022-3, and the Fundação de Amparo à pesquisa do Estado do RS (FAPERGS, Research Support Foundation of the State of RS) for partial financial support under the project No. 23/2551-0000848-3.  J.G.C. is grateful for the support of FAPES (1020/2022, 1081/2022, 976/2022, 332/2023) and CNPq (311758/2021-5, 306018/2025-0). J.C.N.A. thanks FAPESP (2013/26258-4) and CNPq (308367/2019-7) for partial financial support. Some data analyses in this work used the \hyperlink{http://gppd-hpc.inf.ufrgs.br}{PCAD infrastructure} at INF/UFRGS.
\end{acknowledgments}

\bibliographystyle{apsrev4-1}
%

\end{document}